\documentclass[aps,prl,showpacs,twocolumn,amsmath,amssymb,floatfix]{revtex4-1}
\usepackage{tabularx}
\usepackage{bm}
\usepackage{epsfig,subfigure}
\usepackage{graphicx}
\usepackage{color}
\usepackage{amsfonts}
\usepackage{exscale}
\usepackage{amsbsy}
\usepackage{tikz}
\usetikzlibrary{calc}
 \usepackage[colorlinks,urlcolor=blue,linkcolor=blue,anchorcolor=blue,citecolor=blue]{hyperref}
\begin{document}
\title{Measuring Majorana fermions qubit state and non-Abelian braiding statistics in quenched inhomogeneous spin ladders}
\author{Yin-Chen He$^{1}$, and Yan Chen$^{1,2}$}

\affiliation{$^{1}$Department of Physics, State Key Laboratory of Surface Physics and Laboratory of Advanced Materials, Fudan
University, Shanghai 200433, China \\
$^{2}$ Department of Physics and Center of Theoretical and Computational Physics, The University of Hong Kong, Pokfulam Road,
Hong Kong, China}

\begin{abstract}

We study the Majorana fermions (MFs) in a spin ladder model. We propose and numerically show that the MFs qubit state
can be read out by measuring the fusion excitation in the quenched inhomogeneous spin ladders. Moreover, we construct an exactly solvable T-junction spin ladder model,
which can be used to implement braiding operations of MFs. With the braiding processes simulated numerically as non-equilibrium quench processes,
we verify that the MFs in our spin ladder model obey the non-Abelian braiding statistics. Our scheme not only
provides a promising platform to study the exotic properties of MFs, but also has broad range of applications in topological quantum computation.
\end{abstract}


\maketitle
Majorana fermions (MFs) are self-conjugate quasiparticles ($\gamma^{\dagger}=\gamma$)~\cite{Majorana},
and non-Abelian anyons obeying exotic braiding statistics~\cite{braiding,manipulate-kitaevwire}.
Recent years have seen much excitement over MFs, not only because of their peculiar properties, but also due to the possible  applications for
topological quantum computation~\cite{top-comp}. Creating, manipulating and detecting MFs experimentally remain a great challenge, although many theoretical schemes for that have been proposed \cite{FQH,pwave,TI,spin-orb,
kitaev,interferometer1,interferometer2,PhysRevLett.98.010506,qubit-read,Phys.Rev.Lett.98.237002,Phys.Rev.Lett.100.027001,Phys.Rev.Lett.102.216404,Phys.Rev.Lett.103.107002,Phys.Rev.Lett.103.237001, MFs-wire-UCA, SLZhu-2011,Phys.Rev.Lett.106.090503}.
As far as their realization is concerned, an encouraging progress has been made for one dimensional (1D) systems, especially for semiconducting wires~\cite{
superconductingwire}, where a zero-bias conductance peak (ZBCP)~\cite{measure-conductivity} and a fractional Josephson effect~\cite{measure-Josephson} have been
recently measured.
However,  it still remains controversial whether those experimental signatures have shown the realization of MFs \cite{zero-bias, normal-fractional-Josephson}.
To our knowledge, so far there exists no unambiguous and straightforward evidence to demonstrate the braiding statistics of MFs.

Although the superconducting system is a natural choice for the realization of MFs, MFs in spin system receives considerable interest since the pioneer work of Kitaev \cite{kitaev-honeycomb}.
For the realization of MFs in 1D, the spin system is quite different from the electronic system. For instance
the superconductivity can't emerge spontaneously in semiconductor wire, instead it is induced by proximity to a superconductor. This fact imposes extra
difficulty in realization and control of MFs in such system. In contrast, one only needs to engineer the desired spin-spin interaction in the spin system.
Such spin systems may be realized in highly controllable quantum simulation experiments \cite{kitaev-interaction-LMD, spin-interaction-HW, kitaev-interaction-JQY,
kitaev-interaction-RS, kitaev-interaction-SRM}, which may provide promising platform for realizing and controlling MFs.

Despite of recent extensive theoretical studies of MFs in 1D spin system  \cite{spinchain, spinchain2, spinladder1, spinladder2}, it is still unclear about the validity of non-Abelian braiding statistics of MFs
in such system and the way to implement the braiding operations. In this Letter, we propose and
numerically confirm that, one can read out the MFs qubit state by measuring the fusion excitation in spin ladder system with the suppression of KZM excitations by inhomogeneity.
Moreover, we design an exactly solvable T-junction spin ladder model which can be used to implement braiding process. By numerically simulating the braiding operation as a
non-equilibrium process, we verify that the MFs obey non-Abelian braiding statistics, which
provides an avenue for the experimental realization of topological quantum computation.

\begin{figure}[h]
\centering
\includegraphics[width=0.36\textwidth]{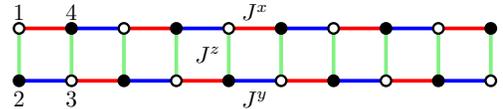}
\caption{\label{fig:ladder} (color online). Spin ladders with three classes of links.}
\end{figure}

\emph{The Hamiltonian and its features.--}
The Hamiltonian of single spin ladder~\cite{spinladder1, spinladder2} can be written as
\begin{equation}
 H=-\sum_{\langle i,j\rangle} J_{ij}^{\beta_{ij}} \sigma_i^{\beta_{ij}}\sigma_j^{\beta_{ij}}, \quad
 \quad \beta_{ij}=x, y,z.\label{eq:Hamiltonian}
\end{equation}
where $\sigma^{x(y,z)}$ are the Pauli operators. We have decomposed the links of the spin ladders into three classes as shown
in Fig. \ref{fig:ladder}, each class of links are associated with one component of
interaction $J^\tau\sigma_i^\tau\sigma_j^\tau$, $\tau=x,y,z$.

To solve the Hamiltonian Eq. (\ref{eq:Hamiltonian}), we represent each spin operator to the product of two MF operators \cite{kitaev-honeycomb, spinladder2}:
\begin{equation}
 \sigma^x_n=ib^x_nc_n \quad\quad \sigma^y_n=ib^y_nc_n \quad\quad \sigma^z_n=ib^z_nc_n \label{eq:Majorana_representation}
\end{equation}
where all the MF operators  $b^{x,y,z}_n$ and $c_n$ satisfy self-adjoint and anti-commutation relations. Using the MF representation,
the Hamiltonian takes the form:
\begin{equation}
 H=\sum_{\langle i,j\rangle} J_{ij}^{\beta_{ij}} (ib_i^{\beta_{ij}}b_j^{\beta_{ij}})(ic_ic_j). \label{eq:Hamiltonian_Majorana}
\end{equation}
It should be noted that, the relation $\sigma_x\sigma_y\sigma_z=i$ imposes a constraint on the product of all MF operators $D_i=b_i^xb_i^yb_i^zc_i=1$. Therefore, after
obtaining the eigenstate $|\psi\rangle$ of the fermionic Hamiltonian Eq. (\ref{eq:Hamiltonian_Majorana}), one should project it into physical Hilbert space:
\begin{equation}
 |\psi\rangle_{phy}=\hat P|\psi\rangle=
 \left(\prod\frac{1+D_i}{2}\right)|\psi\rangle
\end{equation}

In the spin ladder, each site has three links with different interaction terms.
Therefore $u_{ij}=ib_i^{\beta_{ij}}b_j^{\beta_{ij}}$ commutes with Eq. (\ref{eq:Hamiltonian_Majorana}), taking the value $\pm 1$ and as a result we
obtain an exactly solvable quadratic Hamiltonian.
It is convenient to introduce fermionic operator $f_n=(c_{2n-1}+ic_{2n})/2$. Then the Hamiltonian becomes
\begin{align}
 H=\sum_{n=1}^{N-1} &\left[(\omega_{n} f^\dag_{n} f_{n+1}+\Delta_{n}f_{n}f_{n+1}+h.c.)\right. \nonumber \\ &\left. +\mu_{n}(2f_{n}^\dag f_{n}-1)\right]
 \label{eq:p_wave}
\end{align}
where $\omega_{n}=J_{n}^{x}u_{2n-1,2n+2}-J_{n}^{y}u_{2n,2n+1}$, $\Delta_{n}=J_{n}^{x}u_{2n-1,2n+2}+J_{n}^{y}u_{2n,2n+1}$,
$\mu_{n}=J_{n}^{z}u_{2n-1,2n}$. This fermionic Hamiltonian Eq. (\ref{eq:p_wave}) describes Kitaev's p-wave superconducting wire \cite{kitaev}.
For simplicity, we assume that intra-chain coupling $J_n^x,J_n^y$ is positive and homogeneous. Then one can verify that if  $|J_n^z|<|J^x-J^y|$
the groundstate corresponds to all $u_{ij}=1, (i<j)$ \cite{spinladder2},
and such chain is in a topological phase with two MFs located at the end of the chain.

\emph{Measuring MFs qubit state.--}
Two MFs may fuse into either vacuum $|0\rangle$ or one fermion state $|1\rangle$, which can be treated as a qubit state \cite{top-comp}. It is important to measure the
qubit state of two MFs in the process of topological quantum computation as well as experimental realization of MFs. Since MFs are zero energy modes in the topological
phase, both qubit states $|0\rangle$ and $|1\rangle$ are the groundstates of the system, which are hard to distinguish. In electronic system, anyon interference device like
Fabry-Perot interferometer has been designed to detect the MFs qubit state \cite{interferometer1,interferometer2}.
As far as our system is concerned, similar interference device has not yet been invented.

Here we propose a straightforward scheme to realize MFs qubit readout, where one directly fuses the MFs adiabatically and measures the emergent excitation.
The so-called fusing MFs, simply drives the system across the quantum critical point (QCP), from a topological phase to a non-topological phase.
 Then these two MFs in the qubit state $|1\rangle$ will fuse into an excitation, making it easier to detect. Especially, in the spin system, the MFs fusion excitation behaves like a spin flip, which can be easily measured in highly controllable quantum simulation experiment.
However, the danger is that the described process may not be adiabatic due to the vanishing energy gap and divergent relaxation time at the QCP.
These factors inevitably lead to creation of many excitations which number is
determined by the Kibble-Zurek mechanism (KZM) \cite{KZM,KZMreview}.
The KZM excitations may obscure detection of the MFs fusion excitation and therefore
it is necessary to suppress them.
In the following, we introduce inhomogeneity to realize the suppression of KZM excitations~\cite{inhomo}.
This physical result can be understood qualitatively as follows: when an inhomogeneous system undergoes a quench process,
the critical point will be crossed locally and the whole energy spectrum always has finite gap during the quench process.
Moreover, MFs may locate at the natural topological trivial and nontrivial interface yielding by inhomogeneous potential and move
together with critical point during the quench process, which provides a way to manipulate the MFs.

\emph{Numerical simulations.--}
To confirm that one can measure the MFs qubit by the emergent excitations after a quench, we consider two simple processes,
both of which have MFs created and
fused. During the ramp, the inter-chain coupling $J^z$ is inhomogeneous and varies with time:
\begin{align}
\textrm{Process I:} \quad \, \, J_n^z(t) &=\alpha^2 n^2+J_0+t/t_Q\label{eq:potential}  \\ \textrm{Process II:} \quad J_n^z (t)&=\alpha^2 (n-N/2-1/2)^2+J_0+t/t_Q \nonumber
\end{align}
where $\alpha$ denotes the coefficient of parabolic inhomogeneity. In the following, we choose the system size $N=100$.
With an increase of $\alpha$, the minimum gap during the whole adiabatic quench process also increases, making it easier to suppress the KZM excitations.
The term $t/t_Q$ in Eq. (\ref{eq:potential}) represents the quench, with $t_Q$ being  the quench time,
which determines the rate of change in the coupling strength. During the two quench processes, the coupling strength will be ramped from $J_n^z<-|J^x-J^y|$ to
$J_n^z>|J^x-J^y|$.

\begin{figure}[h]%
\centering
\includegraphics[width=0.48\textwidth]{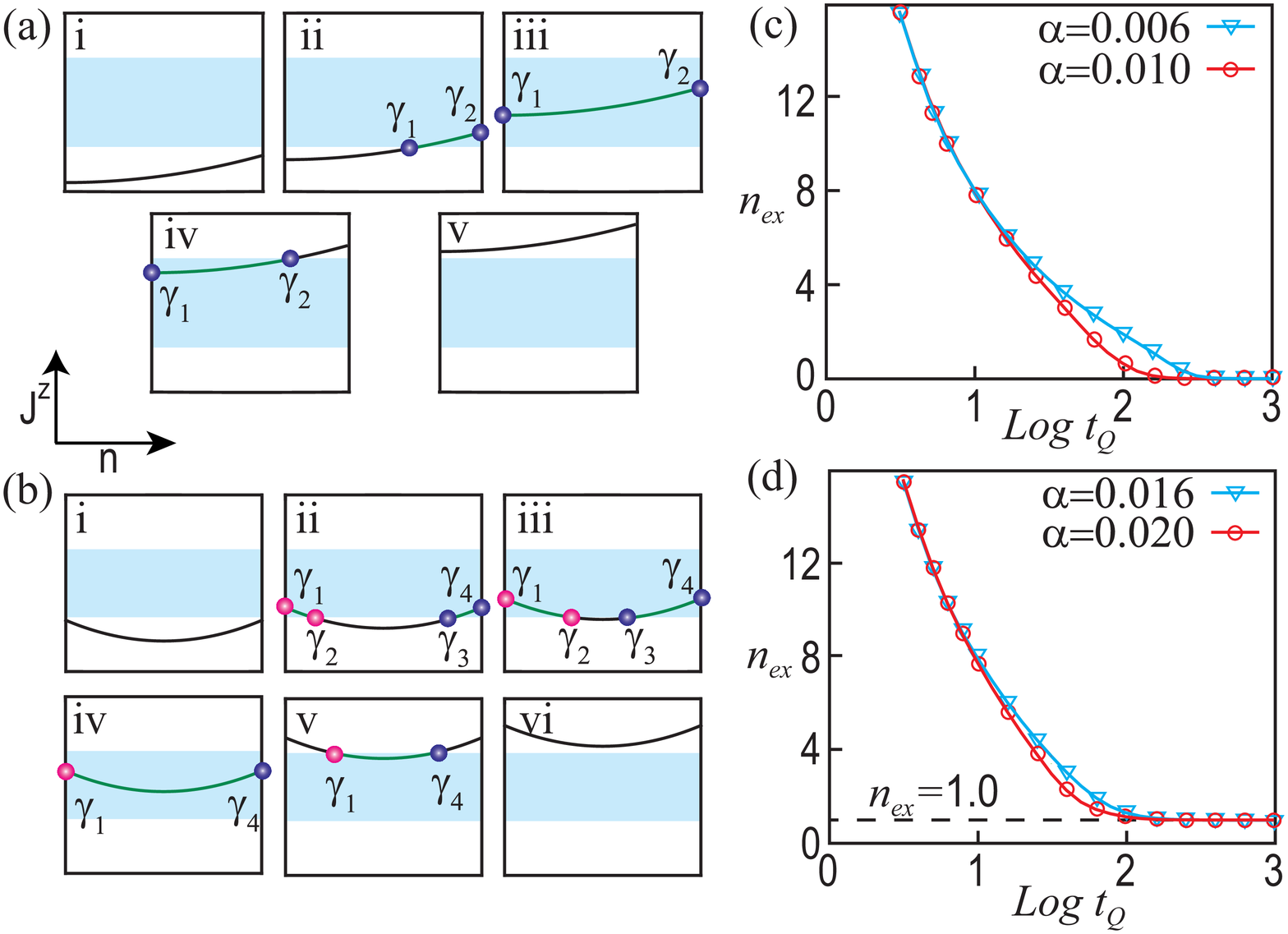}\caption{\label{fig:two_process} (color online). A cartoon picture for the two quench
processes I (a) and II (b). Topological phase regions are marked in light cyan. MFs with the same color, $(\gamma_1,\gamma_2)$ and $(\gamma_3,\gamma_4)$,
are paired MFs, which will fuse into the vacuum state $|0\rangle$.
The number of excitations after the quench process I (c) and II (d). Here we choose $J^x=1.1, J^y=0.1$.}
\end{figure}

These two processes are shown schematically in Fig. \ref{fig:two_process}(a)-(b).
In the quench process I, we create two paired MFs by pulling
them out of vacuum (Fig. \ref{fig:two_process}(a)i-ii). Without participation of other MFs, these two MFs are always in the state $|0\rangle$.
Thus, as one fuses the two MFs (Fig. \ref{fig:two_process}(a)iv-v) to read out the qubit state, no excitations may emerge.
The quench process II has four MFs $(\gamma_1,\gamma_2)$ and $(\gamma_3,\gamma_4)$ pulled out of vacuum (Fig. \ref{fig:two_process}(b)i-ii), whose qubit
state can be written as
$|0,0\rangle$ in the basis $f_A=\gamma_1+i\gamma_2$ and $f_B=\gamma_3+i\gamma_4$.
Interestingly, the two unpaired MFs $\gamma_2$ and $\gamma_3$ will fuse at potential center, as displayed in Fig. \ref{fig:two_process}(b)iii-iv. Therefore, this process
actually measures the MFs qubit state in the basis $f'_A=\gamma_2+i\gamma_3, f'_B=\gamma_1+i\gamma_4$. Written in the $f'_A, f'_B$ basis, $|0,0\rangle$ will be
$(|0'0'\rangle-i|1'1'\rangle)/\sqrt 2$, which implies the emergence of $1$ excitation. It is clear that this read out scheme is insensitive to the relative phase factor
between $|0'0'\rangle$ and $|1'1'\rangle$. However, this relative phase can be read out by fusing the MFs in other pairs, such as $(\gamma_1,\gamma_2)$ and $(\gamma_3,
\gamma_4)$. Generally speaking, for a state ($|0'0'\rangle+ie^{i\theta}|1'1'\rangle)/\sqrt 2$, if one measures it in the basis $f_A=\gamma_1+i\gamma_2, f_B=\gamma_3+i\gamma_4$,
the fusion excitation with number $n=1+\cos\theta$ may emerge.

To verify the above scenario, we perform numerical simulations on the quench dynamics of the two processes. Using the Bogoliubov transformation $a_m^\dag=\sum_n
(u_{nm} f^\dag_n+v_{nm}f_n)$, we can diagonalize the Hamiltonian for any given time, and find that for both quench processes there is always an energy gap \cite{footnote0}.
We consider the BCS groundstate of the ladder $|\psi\rangle=\hat P\prod a^\dag_i|0\rangle/N_0$, and the state after the evolution is
$|\psi_f\rangle=\hat P\prod a^\dag_i(t_f)|0\rangle/N_0$ with
\begin{equation}
 \hat a_m^\dag(t_f)=U \hat a^\dag_m U^\dag, \ \  U=T\left\{\exp\left[-i\int_{t_0}^{t_f} H(t) dt\right]\right\} \label{eq:evolution}
\end{equation}
where $N_0$ is normalization constant, $U(t)$ is time evolution operator, $T$ is time ordering operator.
We also diagonalize the final Hamiltonian $H(t_f)=\sum_m(E_m g^\dag_m g_m-E_m g_m g^\dag_m)$, with $E_m<0$. Then the  number of excitations can be written as
\begin{equation}
 n_{ex}=\sum_m \langle \psi_f| g_m g^\dag_m|\psi_f\rangle. \label{eq:excitation}
\end{equation}

Fig. \ref{fig:two_process}(c)-(d) illustrate the number of excitations for two quench processes, respectively.
It is clear that KZM excitations will be greatly suppressed for long quench time $t_Q$.
In particular, process I gives rise to no excitations while process II yields excitations with universal number 1, which agrees with our previous analysis.

For the general case, the excitation in the spin model may exhibit complex spin configuration,
which may be difficult to detect exactly in experiments. To
clarify the excitations in spin ladder unambiguously,
one can drive the system into  the Ising limit ($J_n^z \gg J^x, J^y$). In this limit, the groundstate shows the parallel alignment of each pair of rung spins from two chains ($\langle J_{2n-1}^zJ_{2n}^z\rangle=1$),
while the lowest excitation corresponds to the anti-parallel alignment of single pair of rung spins ($\langle J_{2n-1}^zJ_{2n}^z\rangle=-1$).
This excitation can
be practically measured in experiments. Therefore, it is rather clear that MFs qubit state can be experimentally read out
through measuring the excitations emerged in quenched inhomogeneous spin ladder.

\begin{figure}[h]
\centering
\includegraphics[width=0.3\textwidth]{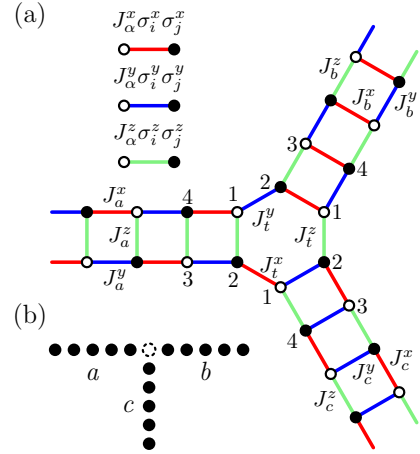}
\caption{\label{fig:model} (color online). (a) Tri-junction spin ladders. Three spin ladders (\emph{a, b, c}) intersect at a hexagon (t). (b) The effective T-junction of Kitaev's superconducting
wire. Compared with the model in Ref. \cite{manipulate-kitaevwire}, the central site at T-junction (white dotted site) is missing in our model.}
\end{figure}

\emph{Non-Abelian braiding statistics--}
First, we design a spin ladder model with T-junction structure to implement the braiding operation of MFs.
Our model is composed of three ladders, which intersect at a hexagon as shown in Fig. \ref{fig:model}(a).

A similar T-junction design for braiding MFs in spin ladders has been proposed \cite{spinladder2}. However, their T-junction model can't be solved exactly, which makes
unclear the fate of the MFs non-Abelian braiding statistics in their model. In contrast,
one may note that with our special design of three ladders and hexagon-junction structure, each site has always three different links. Therefore, this model can be
solved exactly using the Majorana fermionization technique with all the $u_{ij}$ commuting with the Hamiltonian. Following the same procedures before, we obtain fermionic
Hamiltonian $H=\sum_{\alpha=a,b,c} H_\alpha+H_t$, with:
\begin{align}
 H_\alpha&=\sum_{n=1}^{N-1} \left[(\omega_{\alpha,n} f^\dag_{\alpha,n} f_{\alpha,n+1}+\Delta_{\alpha,n}f_{\alpha,n}f_{\alpha,n+1}+h.c.)\right. \nonumber
 \\ &\left.+\mu_{\alpha,n}(2f_{\alpha,n}^\dag f_{\alpha,n}-1)\right],  \label{eq:t_p_wave}
\end{align}
and,
\begin{equation}
  H_t=\sum A_{\alpha\beta} (f^\dag_{\alpha,1} f_{\beta,1}+f_{\alpha,1} f_{\beta,1}+h.c.),
\end{equation}
where $\mu_{a(b,c),n}=2J_{a(b,c),n}^{z(x,y)}u^{a(b,c)}_{2n+1,2n+2}$, $A_{ab(bc,ca)}=J_t^{y(z,x)}u_{ab(bc,ca)}$ and
$\Delta_{a(b,c),n}(\omega_{a(b,c),n})=J_{a(b,c),n}^{x(y,z)}u^{a(b,c)}_{2n+1,2n+4}\pm J_{a(b,c),n}^{y(z,x)}u^{a(b,c)}_{2n+2,2n+3}$. Similar as
before, with $J_{a(b,c),n}^{x(y,z)}$, $J_{a(b,c),n}^{y(z,x)}$, $J_t^{y(z,x)}$ positive, the groundstate corresponds to all $u_{ij}=1$.

The T-junction spin ladder has been mapped into $p$-wave superconducting Kitaev's wire with a T-junction structure, as shown in Fig.  \ref{fig:model}(b). This effective
T-junction Kitaev's wire differs slightly from the model in Ref. \cite{manipulate-kitaevwire}. In particular, the phase of pairing term in our spin model can't take
complex values, its phase can only be $0$ or $\pi$, the change being achieved by adjusting the relative value of intra-chain coupling $J_{a(b,c)}^{x(y,z)},J_{a(b,c)}^{y(z,x)}$.
To realize the braiding of MFs, it is important that in the braiding process, no additional MFs appear at the T-junction. Therefore, the phase of pairing term should
be different for the wire-pairs (\emph{a,c}) and (\emph{b,c}) \cite{manipulate-kitaevwire}. We can choose the pairing phase of wire \emph{a} and \emph{b} to be $0$, while that of wire \emph{c} to be $\pi$. To achieve this we can
simply put $J^x_a>J^y_a, J^y_b>J^z_b$ and $J^z_c<J^x_c$.

\begin{figure}[h]
\centering
\includegraphics[width=0.4\textwidth]{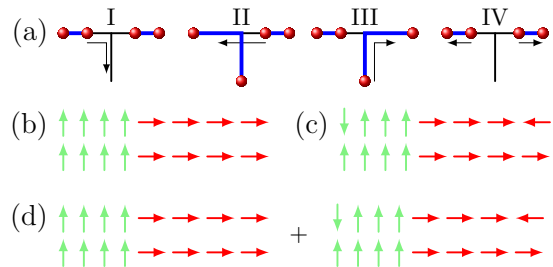}
 \caption{\label{fig:state} (color online). (a) Procedures of braiding and reading out the MFs qubit state.
Schematic representation for the quantum state under the large inter-chain coupling limit, here we omit the ladder \emph{c} for simplicity.
(b) Groundstate of the system: spins on the ladder \emph{a} (\emph{b}) align parallel along $z$ ($x$) direction. (c)
 Output state by braiding MFs twice: excited state with two excitations located at the left and right end. (d) Output state by braiding MFs one (three) times: equally weighted superposition of
two states (b) and (c).}
\end{figure}

To verify unambiguously the braiding statistics, we perform numerical simulations on the non-equilibrium processes which have the MFs braiding counterclockwise, as illustrated
in Fig. \ref{fig:state}(a)I-III. After braiding finite (1-4) times, we perform the qubit read-out procedure by driving the whole system non-topological and then
measuring the emergent fusion excitations, as shown in Fig. \ref{fig:state}(a)IV.
We begin with four MFs ($\gamma_1,\gamma_2$) and ($\gamma_3,\gamma_4$), whose state can be written as $|00\rangle$ in the basis of $f_A=\gamma_1+i\gamma_2$ and
$f_B=\gamma_3+i\gamma_4$.
Apparently, our measurement scheme will read out the qubit state in the basis of $f_A$ and $f_B$. The braiding of MFs $\gamma_2$ and $\gamma_3$
can be described by the unitary operator $U=\exp(\pi\gamma_2\gamma_3/4)$.
By MFs braiding one or three times, the qubit state will be $(|00\rangle\pm |11\rangle)/\sqrt 2$. By MFs braiding twice, the qubit state will be $|11\rangle$.
Furthermore, the qubit state will be changed back into the initial state $|00\rangle$ by MFs braiding four times. Our numerical calculations confirm those results \cite{footnote0}. In particular, we have found the emergence of 1 or 2 excitations after the process with MFs braiding one (three) or two times, with each half of the excitations emerged at the left (right) end of \emph{a} (\emph{b}) ladder. At last, braiding MFs four times produces no excitations.

The general scheme to illustrate the non-Abelian braiding statistics can be visualized as follows. First we prepare a non-topological state, which is
schematically shown as in Fig. \ref{fig:state}(b). To create and manipulate the MFs, one only needs to simply change the inter-chain coupling strengths of three ladders.
After braiding the MFs one or three times and driving the system back into non-topological phase, one may obtain
a quantum state as shown in Fig. \ref{fig:state}(d). After braiding the MFs twice, one may obtain two excitations localized at the ends of
spin ladders \emph{a} and \emph{b}, as shown in Fig. \ref{fig:state}(c). Finally, one may perform the braiding of the MFs four times, and the system
goes back to its groundstate.

\emph{Discussion.}--
It is promising to realize topological quantum computation using a network of our T-junction spin ladders. Nevertheless, there are two related
outstanding theoretical issues.
First, our present qubit read-out scheme is still destructive and a non-destructive qubit read-out scheme \cite{PhysRevLett.98.010506, qubit-read}
is definitely demanded.
With highly controllable quantum simulation techniques, one can practically conduct
non-destructive measurement to detect the quantum states of our proposed spin system, including the MFs qubit state.
Meanwhile, with a better qubit read-out scheme, the requirement of adiabaticity may become unnecessary,
which may further facilitate the experimental realization.
Second, as we know, the braiding operations of MFs are unable to realize all the quantum gates necessary for universal quantum computation.
One possible way to achieve other necessary quantum gates is to generate an effective interaction between the unpaired MFs,
and the resultant MFs tunneling may serve as quantum gates. The delicate control of the
interaction between MFs may realize the desired quantum gates. An alternative way is to construct a spin model whose
dual fermionic superconducting Hamiltonian has complex phase factor in the pairing term, then a phase gate \cite{top-comp} may be implemented.

\emph{Summary.}--
We have proposed and numerically shown that, in a quenched inhomogeneous spin ladder model, one can read out the MFs qubit state by measuring the MFs fusion
excitations. An exactly solvable T-junction
spin ladder model is designed to implement MFs braiding operation. With numerical simulation on the non-equilibrium braiding process,
we show that the MFs in our model obey non-Abelian braiding statistics. Our proposal may be realized in quantum simulation experimental system like
ultracold atoms or Josephson junctions, which may pave the way for the realization of the topological quantum computation using 1D spin system.

We acknowledge J.Q. You and Y. Yu for useful discussions. Y.-C. He appreciates the help of Y.-Y. Zhao, K. Ding and J.-D. Zang.
This work was supported by the State Key Programs of China
(Grant Nos. 2012CB921604 and 2009CB929204) and the National Natural Science Foundation
of China (Grant Nos. 11074043 and 11274069) and Shanghai Municipal Government,
the RGC grants in HKSAR.

\pagebreak

\onecolumngrid

\begin{center}
{\bf \large Supplementary material}
\end{center}
\section{Energy spectrum}
\begin{figure}[h]
 \centering
 \includegraphics[width=0.6\textwidth]{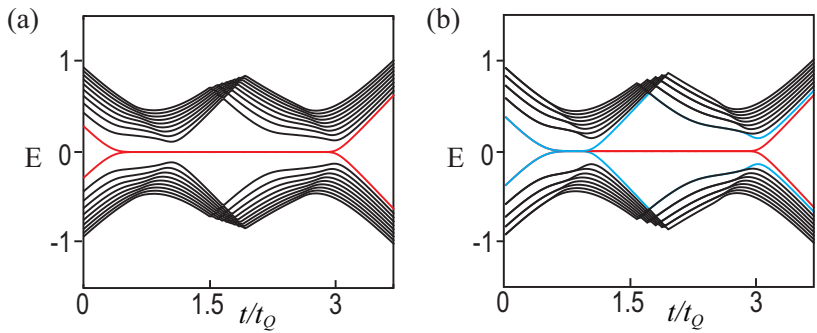}\caption{Energy spectrum of quench process I (a) and II (b). In (a), we take the parameters as $\alpha=0.01$,
$h_0=-2.1$, $N=100$, and $t/t_Q$ from $0$ to $3.7$, $\omega=1, \Delta=1.2$. In (b), we take the parameters as $\alpha=0.02$, $h_0=-2.1$, $N=100$, and $t/t_Q$ form $0$
to $3.7$, $\omega=1, \Delta=1.2$. For simplicity, we only plot several states with energy around $0$. \label{fig:sup_energy}}
\end{figure}

Using a Bogoliubov transformation $a_m^\dag=\sum_n (u_{nm} f^\dag_n+v_{nm}f_n)$, Kitaev's wire Hamiltonian
\begin{equation}
 H=\sum_{n=1}^{N-1} \left[(\omega f^\dag_{n} f_{n+1}+\Delta f_{n}f_{n+1}+h.c.) +\mu_{n}(2f_{n}^\dag f_{n}-1)\right]
 \label{eq:sub_p_wave}
\end{equation}
can be diagonalized into $H=\sum_m (E_m a_m^\dag a_m-E_m a_m a^\dag_m)$, where $E_m<0$. Due to the particle-hole symmetry,
the spectrum has conjugate pairs with negative/positive energy, and we call $a^\dag_m$ quasiparticle, $a_m$ quasihole.
One can obtain the energy spectrum during the two quench processes:
\begin{align}
\textrm{Process I:} \quad \, \, J_n^z(t) &=\alpha^2 n^2+J_0+t/t_Q\label{eq:sub_potential}  \\ \textrm{Process II:} \quad J_n^z (t)&=\alpha^2 (n-N/2-1/2)^2+J_0+t/t_Q \nonumber
\end{align}
The energy spectrum is shown in Fig. \ref{fig:sup_energy}. One can observe that the there is always an energy gap during the quench process, and Majorana zero modes will appear when part of the chain is engineered into topological phase. Especially, for the
process II, four Majorana fermions (MFs) will appear.

 \section{Number of emergent excitations}
For the Kitaev's superconducting wire, the eigenstate can be constructed by $N$ particles, each of them is chosen from $(a^\dag_m,a_m)$ conjugate pairs, which reads,
\begin{equation}
|\psi\rangle=\prod_{j\in A}a_j\prod_{i\in N-A} a_i^\dag|0\rangle/N_0 \label{eq:sub_BCS}
\end{equation}
where $N_0$ is the normalization constant. To calculate the number of excitations created by the quench process,
we start with a groundstate, $|\psi_0\rangle=\prod_i a_i^\dag |0\rangle/N_0$, evolve the state into $|\psi_f\rangle$, with time-evolution operator $U=T\left\{\exp\left[-i\int_{t_0}^{t_f}
H(t) dt\right]\right\}$:
\begin{equation}
|\psi_f\rangle= U|\psi_0\rangle=U\prod_i a_i^\dag |0\rangle/N_0=\prod_i (Ua_i^\dag U^\dag)|0\rangle \label{eq:sup_evol}
\end{equation}

Therefore, we can simply calculate the time-evolution of each quasiparticle $a_m^\dag$, $a^\dag_m(t_f)=Ua^\dag_m U^\dag$. Apparently, the evolved quasiparticle $a_m^\dag(t_f)$ still satisfy the fermionic commutation relation $\{ a_m(t_f), a_n(t_f)\}=
\{ a_m^\dag(t_f), a_n^\dag(t_f)\}=0$, and $\{ a_m^\dag(t_f), a_n(t_f)\}=\delta_{mn}$.

To implement the time-evolution numerically, one may discretize the time-evolution operator by using the time-slicing procedure:
\begin{equation}
 U=T\left\{\exp\left[-i\int_{t_0}^{t_f} H(t) dt\right]\right\}\approx \prod_t \exp\left[ -i H(t_i)\Delta t\right] \label{eq:sup_evol2}
\end{equation}
with $\Delta t\ll 1$. It should be noted that it is crucial to retain the unitarity of $\exp\left[ -i H(t_i)\Delta
t\right]$ throughout the calculation:
\begin{equation}
\exp\left[ -i H(t_i)\Delta t\right]=A \exp(- i \Lambda \Delta t) A^\dag \label{eq:sup_evol3}
\end{equation}
where $ H(t_i) =A \Lambda A^\dag$, $A$ is a unitary matrix $AA^\dag=I$ and $\Lambda$ is a diagonal matrix.

Meanwhile, one can diagonalize the final Hamiltonian $H(t_f)$, obtains Bogoliubov quasiparticles $(g_1^\dag,\cdots,g_N^\dag)$,
and Bogoliubov quasiholes $(g_1,\cdots, g_N)$. The groundstate of the final Hamiltonian is composed with all the quasiparticles
$(g_1^\dag,\cdots,g_N^\dag)$. Therefore, the excitation number in the evolved state $|\psi_f\rangle$ is $n_{ex}=\sum_i \langle \psi_f| g_i g_i^\dag |\psi_f \rangle$.
We may rewrite $b_n$ as superposition of $a^\dag_m(t_f)$ and $a_m(t_f)$:
\begin{equation}
 g^\dag_n=\sum_m [\beta_{nm}^* a_m^\dag(t_f)
+\eta_{nm} a_m(t_f)].
\end{equation}
The excitation number can be written as
\begin{equation}
 n_{ex}=\sum_{n,m}|\eta_{nm}|^2
\end{equation}

For the spin-ladder system, one should note that it is necessary to project the BCS state into the physical Hilbert space by operator $\hat P$
\begin{equation}
|\psi\rangle_{phy}= \hat P|\psi\rangle=\prod\left( \frac{1+D_i}{2}\right)|\psi\rangle.
\end{equation}
One may be worried that this projection will bring some complexity here. However, the projection operator $\hat P$ commutes with the Hamiltonian, as result of which,
all the calculation about the non-equilibrium dynamics of Kitaev's wire is also valid for spin-ladder with projection.  

\section{Numerical Results of Braiding operations}
To have MFs braided, one should tune the inhomogeneous inter-chain coupling $J_n^z$, so as to move the MFs hosted at the interface between topological ($J_n^z
<|J^x-J^y|$) and non-topological ($J_n^z>|J^x-J^y|$). This is a non-equilibrium process which can be calculated using Eq. (\ref{eq:sup_evol}), (\ref{eq:sup_evol2})
and (\ref{eq:sup_evol3}). 

After accomplishing the braiding process, we read out the qubit state by driving the system topological and tune the inter-chain coupling $J^z_n$ to limit $J^z_n
\gg J^x_n, J^y_n$. In this limit, the groundstate corresponds to all $\langle J^z_{2n-1}J^z_{2n}\rangle=1$, while the $\langle J^z_{2n-1}J^z_{2n}\rangle=-1$ is a
single excitation.

We carry out the numerical simulation on such non-equilibrium process, obtain that the number of emergent excitations is universally 1 or 2 for MFs braiding one (three)
or two times. For MFs braiding four times, there emerges on excitation. We plot the excitation distribution ($\langle J^\beta_{2n-1}J^\beta_{2n}\rangle$) in Fig.
{\ref{fig:braiding_numerical} (a)-(b), and one may find that the excitations are mainly localized at the left and right ends \cite{sup_footnote1}.

To facilitate the detection of MFs fusion excitation, we can employ a trap potential at the left (right) end of ladder \emph{a} (\emph{b}), where inter-chain coupling $J_1^z$ ($J_{2n}^x$) \cite{sup_footnote2} at the left (right) end
of ladder \emph{a} (\emph{b}) is set little smaller than the other site. Since the MFs fuses at the left (right) end of ladder \emph{a} (\emph{b}), we can expect that the MFs fusion excitation will be
trapped exactly at the left (right) end of ladder \emph{a} (\emph{b}). 

The numerical results are shown in Fig. \ref{fig:braiding_numerical}(c)-(d). It clearly shows that for MFs braiding one or three times (Fig. \ref{fig:braiding_numerical}(c)),
one will have two half excitations emergent,
each of them is well located at the left (right) end of ladder \emph{a} (\emph{b}). While for MFs braiding twice (Fig. \ref{fig:braiding_numerical}(d)), two excitations emerges,
which are also located the the ends of the ladders. Finally, for MFs braiding four times, we find no excitations emerges.

\begin{figure}[h]
 \centering
 \includegraphics[width=0.7\textwidth]{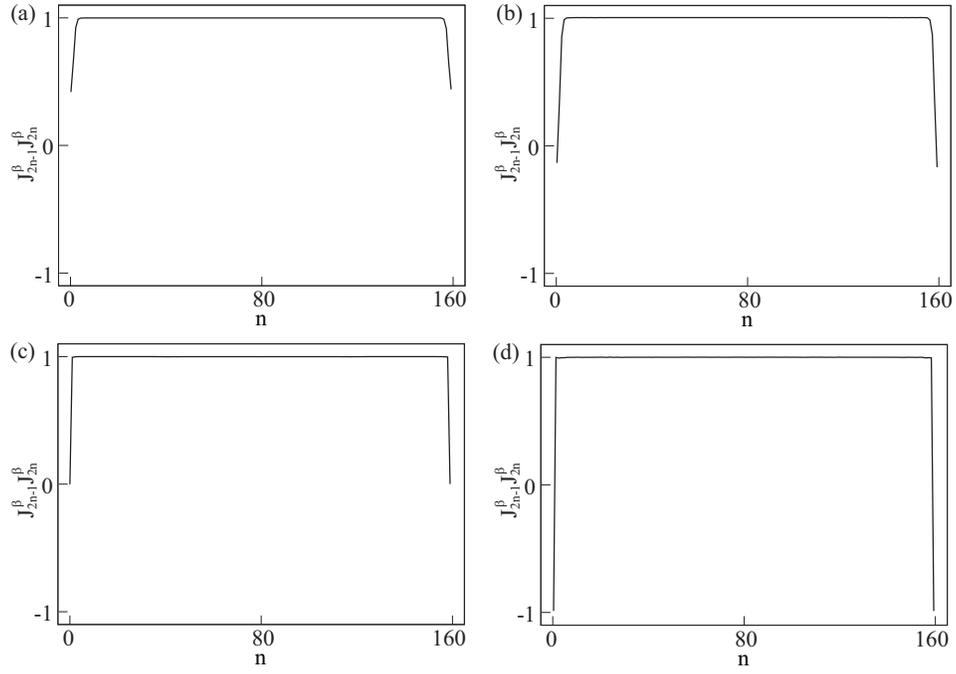}\caption{\label{fig:braiding_numerical} Distribution of excitations after the braiding operations. Here we
 calculate system size with $N=80$ each ladders. For simplicity, we only plot results of ladder $a$ and $b$. For $n=1,\cdots, 80$, the plotted spin-spin correlation is $\langle J_{2n-1}^z
 J_{2n}^z\rangle$, while for $n=81,\cdots, 160$, the correlation is $\langle J_{2n-1}^xJ_{2n}^x\rangle$  (a)
 Braid MFs one or three times without trap potential. (b) Braid MFs two times without trap potential.  (c)
 Braid MFs one or three times with trap potential. (d) Braid MFs two times with trap potential.}
\end{figure}

\end{document}